\documentclass[twocolumn,superscriptaddress,floatfix,showpacs,aps]{revtex4}%
\usepackage{eurosym}
\usepackage{graphicx}
\usepackage{dcolumn}
\usepackage{ulem}
\usepackage{textcomp}
\usepackage{amssymb}
\usepackage{amsmath}
\usepackage{color}
\usepackage{footnote}
\usepackage{microtype}
\usepackage{amsfonts}%
\providecommand{\U}[1]{\protect\rule{.1in}{.1in}}
\providecommand{\U}[1]{\protect\rule{.1in}{.1in}}

\begin{document}
\title{Prediction of unconventional magnetism in
doped FeSb$_{2}$}
\author{I. I. Mazin}
\affiliation{Department of Physics and Astronomy, George Mason University, Fairfax, USA}
\affiliation{Center for Quantum Science and Engineering, George Mason University, Fairfax, USA}
\author{K. Koepernik}
\affiliation{Institute for Theoretical Solid State Physics, IFW Dresden, 01069 Dresden, Germany}
\author{M.D. Johannes}
\affiliation{Center for Computational Materials Science, Naval Research Laboratory,
Washington D.C., USA}
\author{Rafael Gonz\'{a}lez-Hern\'{a}ndez}
\affiliation{Grupo de Investigaci\'{o}n en F\'{\i}sica Aplicada, Departamento de
F\'{\i}sica,Universidad del Norte, Barranquilla, Colombia}
\affiliation{Institut f\"ur Physik, Johannes Gutenberg Universit\"at Mainz, 55128 Mainz, Germany}
\author{Libor \v{S}mejkal}
\affiliation{Institut f\"ur Physik, Johannes Gutenberg Universit\"at Mainz, 55128 Mainz, Germany}
\affiliation{Institute of Physics, Czech Academy of Sciences, Cukrovarnick\'a 10, 162 00,
Praha 6, Czech Republic}
\date{\today }

\begin{abstract}
It is commonly believed that in typical collinear antiferromagnets, with no net magnetization, 
 the energy bands are spin-(Kramers-)degenerate. The opposite case is usually associated with 
a global time-reversal symmetry breaking (e.g., via
ferro(i)magnetism), or with the spin-orbit interaction is combined with the broken
spatial inversion symmetry. Recently, another type of spin splitting was
demonstrated to emerge in some fully compensated by symmetry, {
nonrelativistic, collinear magnets, and not even necessarily non-centrosymmetric}. These materials feature non-zero spin density staggered not only 
in real, but also in momentum space. This duality results in a combination of characteristics typical
of ferro- and antiferromagnets. Here we discuss this novel
concept in application to a well-known semiconductor, FeSb$_{2},$ and predict
that upon certain alloying it becomes magnetic, and features such magnetic duality. The calculated energy bands split
antisymmetrically  {with respect to spin degenerate nodal surfaces (and not 
nodal {\it points}, as in the case of spin-orbit splitting}. This combination of a  large (0.2 eV) spin splitting, 
compensated net magnetization and metallic ground-state, and a particular 
magnetic easy axis generate a large anomalous Hall conductivity ($\sim$150
S/cm) and a sizable magneto-optical Kerr effect, all deemed to be hallmarks of
nonzero net magnetization. We identify a large contribution to the anomalous
response originating from the spin-orbit interaction gapped anti-Kramers nodal
surfaces, a mechanism distinct from the nodal lines and Weyl points in
ferromagnets.

\end{abstract}
\maketitle

Antiferromagnets are commonly associated with spin-degenerate bands throughout
the entire Brillouin zone. The reason is that while antiferromagnets break the
time-reversal symmetry $\mathcal{T}$ only microscopically, they preserve it,
by definition, when combined with another symmetry operation $\mathcal{O}$.
Examples of such operations are a lattice translation $\boldsymbol{t}$
(\textit{i.e.,} doubling of the unit cell) or a spatial inversion
$\mathcal{P}$. In those cases the combined symmetry and
protects the Kramers spin degeneracy for all wavectors.

 Recently, it was
pointed out that this is not necessarily always the case. There exists magnets where
the collinear spin densities (we do not discuss noncollinear spin textures in
this paper) are perfectly compensated but their particular spatially
anisotropy leads to a strong spin splitting of energy
bands\cite{Libor,Ahn2019,Japan,Zunger,Noda2016,Lopez-Moreno2012} and surprisingly strong anomalous
responses such as spontanous\cite{Reichlova2020} and crystal Hall
effect\cite{Libor,Feng2020a}, crystal magneto-optical Kerr
effect\cite{Samanta2020}, spin polarized
currents\cite{Gonzalez-Hernandez2021,Naka2020c} and giant magnetoresistance
effects \cite{Reichlova2020,Smejkal2021,Shao2021}. The material base for this
emerging class of magnets is potentially large, but so far relatively few
materials have been convincingly
identified\cite{Libor,Feng2020a,Reichlova2020} and we only started to explore
their fascinating electronic structure. In addition, many of these systems are
insulating, eliminating or suppressing many of the effects above, or their
ground-state easy axis is incompatible with the anomalous response. 

From the above description it follows that such unconventional magnets
 must involve more than one crystallographically equivalent magnetic site
in the nonmagnetic unit cell (otherwise the operation $\mathcal{O}$ would be a
lattice translation). Furthermore, if the structure includes at least one bond
between two ions with the opposite spins such that its
middle point is an inversion symmetry center, this structure is and ordinary 
antiferromagnet with Kramers degeneracy
(such an operation would map a spin-up states onto a spin-down state with the
same momentum). Incidentally, the same condition is usually invoked regarding
Dzyaloshinskii-Moria interaction. 

One way to break this symmetry is to surround magnetic ions by tilted cages of
nonmagnetic atoms\cite{Libor} such as in the marcasite structure. The arguably best-explored
marcasite material FeSb$_{2}$ is experimentally nonmagnetic, albeit exhibits
unusual properties including a magnetic response that changes from diamagnetic
to paramagnetic. Experimentally, the FeSb$_{2}$ transport switches from weakly
semiconducting at higher temperatures to metallic at $T\sim50-100$ K. The
optical gap has been measured between 76 meV \cite{optics_Homes} and 130 meV
\cite{optics_Herzog} and the transport gap is anisotropic with multiple gaps
between 4 meV and 36 meV \cite{optics_Homes,optics_Herzog,res_Sun}. Various
density functional theory (DFT) approximations \cite{BS_Kang, BS_Tomczak,
BS_Bentien} have found a pseudogap and semiconducting non-magnetic ground
state with an indirect gap.
%Although all transport measurements confirm the semiconducting state,
%quantum oscillation measurements reveal clear Fermi surfaces\cite{Sebastian}.
%Additionally, the Fermi surfaces shifts as a function of field magnitude and
%direction are argued to be the result of a ferromagnetic response strongly
%coupled to the lattice via spin-orbit effects \cite{Sebastian}.
 Doping with
both Cr\cite{Cr} and Co\cite{Co} has been effected, and leads to magnetic
phases, some of them not entirely determined.

In this paper we predict, using first-principles calculations, that the
stoichiometric FeSb$_{2}$ is, on the mean-field level, an ordinary
antiferromagnet with Kramers degenerate bands, with the structure traditionally called AFMe and shown in
Fig.~1(a). However, in our calculations this structure is nearly degenerate with another magnetic
structure, AFMo (Fig.~1(b)), which is, actually, an {un}conventional ``dual'' magnet as 
descibed above. The calculated energy of the nonmagnetic state is also nearly degenerate with the other two. We argue that the
antiferromagnetic state shown in {Fig.~1(a), is suppressed in the
stoichiometric compound due to to spin fluctuations.  Importantly, these energies are close despite the fact 
that the AFMo structure, being a good
metal, is disadvanteged by kinetic energy. Not surprisingly, when this disadvantage is removed by doping FeSb$_{2}$ 
away from the semiconducting gap, the unconventional spin-split AFMo state is stabilized, such as for moderate hole (Cr) or electron (Co) doping.
Larger dopings stabilize the trivial AFMe structure.

The nonrelativistic band-structure of the AFMo phase exhibits spin-split
Fermi surfaces as illustrated in Fig.~1(c), with the nodal planes as shown in
{Fig}. 1(d). The spin splitting follows a higher symmetry than the underlying
orthorhombic crystal structure, since the reflection about these plane changes
the sign, but not the amplitude of the splitting. It can therefore be expanded
into lattice harmonics as
\begin{equation}
E(\boldsymbol{k},\uparrow)-E(\boldsymbol{k,\downarrow})=F(\boldsymbol{k}%
)\sin(k_{x}a)\sin(k_{y}b),\label{d}%
\end{equation}
where $F(\boldsymbol{k})$ respects the full underlying lattice symmetry and $a$, $b$  {are}
 the lattice parameters. We
refer to the resulting symmetrically spin split bandstructure with Kramers
nodal \textit{surfaces} as anti-Kramers (AK) to distinguish it from the
conventional Rashba antisymmetric spin splitting with spin degenerate Kramers
\textit{points}. Furthermore, we show by direct relativistic first-principle
calculations that the spin-orbit interaction gapped AK nodal surfaces
significantly contribute to a large spontaneous crystal Hall conductivity
$\sigma_{xy}\sim150$~S/cm and magnetooptical response on the level of
$\varepsilon_{xy}(\omega)\sim\pm10$. We will show that the AK mechanism
ensures large Berry curvature at Fermi surfaces. This AK mechanism is very robust, in sharp contrast with the
conventional mechanisms operative in ferromagnets, which rely on fine-tuning
of nodal lines or Weyl points close to the Fermi level
\cite{Gosalbez-Martinez2015,Kim2018d,Liu2018c}. Since CrSb$_{2}-$FeSb$_{2}%
-$CoSb$_{2}$ form a continuous solid solution with a marcasite
(pseudo-marcasite for high Co content) structure\cite{Co,Cr}, such doping
should be accessible experimentally and thus this system represents a prime
candidate for testing the predicted anomalous responses.

%\onecolumngrid

\begin{figure}[ptb]
\includegraphics[width =.95 \linewidth] {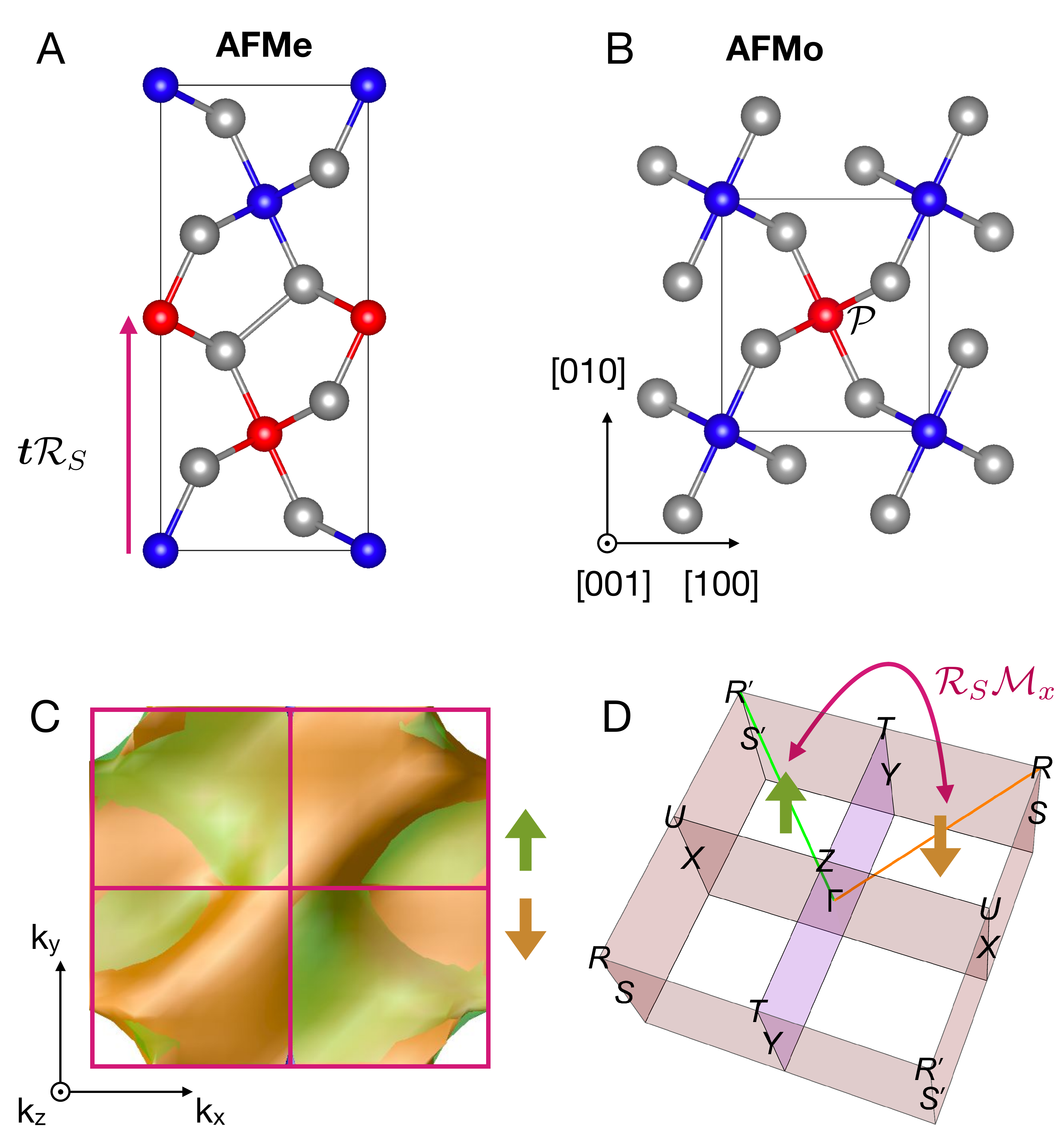}
\caption{\textbf{{Magnetic} ground-states in FeSb$_{2}$.} The Kramers
(a) and spin-split (b) ordering shown on the FeSb$_{2}$
structure. Fe atoms are shown in red ("up" spin) and blue ("down" spin) and Sb
atoms are shown in light grey. In the K state we mark the half-unit cell
translation coupled with time-reversal $\mathbf{t}\mathcal{T}$. In the AFMo
state we mark the inversion symmetry $\mathcal{P}$. (c) Top view along the
c-axis on the spin polarized Fermi surface calculated for the spin split antiferromagnetic FeSb$_{2}$
state. (d) Brillouin zone with marked spin degenerate anti-Kramers nodal surfaces. We
also show the path $\Gamma R$ and $\Gamma R^{\prime}$ to emphasize that it is
orbitally degenerate but with an opposite spin polarization (marked by green
and orange) and we mark the magenta $\mathcal{R}_{S}\mathcal{M}_{x}$ plane. }%
\label{AFMpattern}%
\end{figure}

%\twocolumngrid

\section{Antiferromagnetic ground-states in FeSb$_{2}$}

The marcasite crystal of FeSb$_{2}$ exhibits the nonmagnetic orthorhombic
symmetry space group $Pnnm$. The previous first principle studies of assumed
that the Fe electrons are strongly correlated, resulting in a competitive
meta-ferromagnetic state with Fe magnetization on the order of 1 $\mu_{B}%
$.\cite{Lukoyanov2006} However, these calculations did not address any
possibility of antiferromagnetic order in FeSb$_{2}$. Meanwhile, there is
study of antiferromagnetism of another marcasite crystal, CrSb$_{2}%
$.\cite{DFT}

In our calculations, we consider two antiferromagnetic orderings shown in
Fig.~1(a) and (b). The former, known in the literature as AFMe\cite{DFT}, is an ordinary
Kramers antiferromagnet (KAF). It is obtained by quadrupling the nonmagnetic
unit cell along [100] and [001] axes.
The latter ordering (Fig.~1(b)), known as AFMo\cite{DFT}, preserves the size
of the nonmagnetic unit cell and is an {AK magnet}, where the 
spin polarization is non-zero, and changes sign both in real (from one Fe to the other), and in reciprocal (across the nodal planes) space.

In Fig.~2(a) we show our calculated ground-state energy differences as a
function of doping on the Fe site. We employed both virtual crystal
approximation (VCA) and supercell first-principle calculations (see Methods).
First, we focus on the electron and hole doping around the Fe stoichiometric
point corresponding to the blue curve in Fig.~2(a). The KAF is lower in energy
than the nonmagnetic state (blue curve) in this region except in the vicinity
of the stoichiometric Fe point, where it is only marginally ($\sim1$ meV/Fe)
more stable. Enhancing the tendency to magnetism by adding Hubbard U (LDA+U),
obviously, stabilizes the KAF magnetic solution even further,
albeit, given the experimental situation, LDA+U may be in fact less accurate
than the straight DFT (see Methods for the choice of the density functional used).

Next, we discuss the energetics of the two antiferromagnetic phases (the green
curve in Fig.~2(a)). For the stoichiometric Fe, Cr and MnSb$_{2}$ the KAF
state is lower than the AK one by, for instance, $\approx11$ meV, or
$\approx120$ K per Fe (see Fig.~2(a)) for pure FeSb$_{2}$. The lower energy of
the KA structure can be related to opening a pseudogap in the density of
states, nearly identical to the non-magnetic pseudogap, while the spin split antiferromagnetic phase
is a good metal, with the corresponding loss of the one-electron energy. There
are two corollaries of this finding: first, the near-degeneracy of two rather
different magnetic states, as it is common in frustrated magnetic systems, is
liable to suppress both ordered states and stabilize the dynamically
nonmagnetic one. Second, one can conjecture that as long as the material is
doped away, in either direction, and this disadvantage of the spin split antiferromagnetic phase is
alleviated, the latter will become the most stable. These strong fluctuations
may be playing an important role in thermoelectricity FeSb$_{2},$ through an
energy dependence of the scattering rate, but such an analysis is beyond the
scope of this paper.

As expected, our VCA calculations predicts that moderate electron or hole
doping does stabilize the spin split antiferromagnetic order. To illustrate the role of the pseudogap,
we compare the FeSb$_{2}$ density of states with that for the electron-doped
Co$_{0.3}$Fe$_{0.7}$Sb$_{2}$ in Fig.~2(c) The spin split antiferromagnetic ordering
has the \textit{highest }weight at the Fermi energy in the former, and the
\textit{lowest} in the latter case. Total energy calculations in Fig.~2(a)
confirm that in the calculations the AFMo structure is the most stable in same
range of either electron and hole doping. The energy gain
due to the antiferromagnetism with respect to the nonmagnetic phase grows
extremely rapidly with doping, suggesting that except of the close vicinity to
the stoichiometric compound the competition is between different magnetic phases.

\begin{figure}[tb]
\includegraphics[width = 1\linewidth] {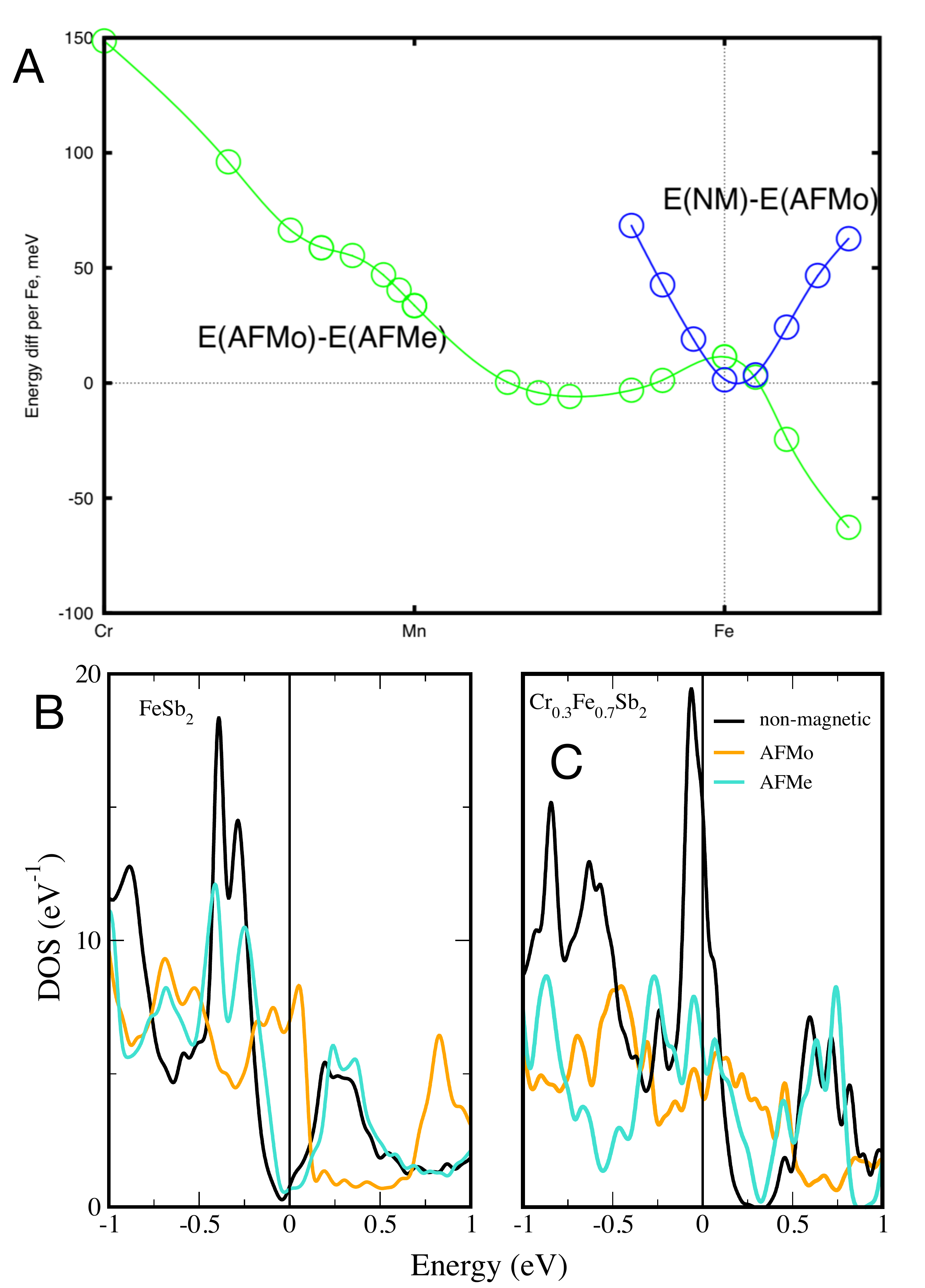}\caption{ (a) A plot of the
total energy of FeSb$_{2}$ with different imposed magnetic orderings as a
function of hole and electron doping in the virtual crystal approximation
(VCA) approximation. The single point at Cr=0.5 corresponds to an actual
Cr-doped calculation with full structural relaxation (non-VCA) and illustrates
that the VCA captures the relevant energy scale. (b, c) Density of states for
FeSb$_{2}$ and Co$_{0.3}$Fe$_{0.7} $Sb$_{2}$ in the VCA, illustrating the
removal of weight from E$_{F}$ upon hole doping in the AFMo and FM patterns,
but an increase in weight for non-magnetic state. The KA gains some weight at
E$_{F}$ }
\label{DOS}%
\end{figure}

\section{Nonrelativistic symmetry analysis of anti-Kramers spin splitting}

We now consider symmetries of the magnetic phases. The spin
degeneracy in the FeSb$_{2}$ KAF phase is due to the combined $\boldsymbol{t}%
\mathcal{T}$ symmetry in the magnetic point group \cite{Smejkal2016}.
Alternatively, one can say that it is protected by the combined (as opposed to
individual) $\mathcal{PT}$ symmetry. The antiferromagnetism is constructed by
quadrupling the unit cell, $i.e.$, doubling along both the [100] and
[001] crystal axes. The opposite magnetic sublattices in the KAF phase are
thus connected by the nonmagnetic unit cell translations combined with
time-reversal. %Since spacial translation does not affect the band dispersion
%in the reciprocal space, the two spin states are degenerate.

Let us now consider the  {unconventional AK} phase. There, magnetic order} does
not change the size of the unit cell and the crystal lacks both
$\boldsymbol{t}\mathcal{T}$ and $\mathcal{PT}$ symmetries, and thus allows for
a spin splitting at a general wave vector.

If we neglect the spin-orbit interaction, the real space and the spin space
are decoupled. The resulting nonrelativistic symmetry space
group\cite{Reichlova2020} of the FeSb$_{2}$ includes the following symmetry
operations:
\begin{equation}
\left\{  \mathcal{E},\mathcal{P},\mathcal{M}_{z},\mathcal{C}_{2z}\right\}
+\boldsymbol{t}^{\prime}\mathcal{R}_{S}\left\{  \mathcal{C}_{2x}%
,\mathcal{M}_{x},\mathcal{C}_{2y},\mathcal{M}_{y}\right\}  .\label{nonrelativ}%
\end{equation}
Here the $\mathcal{R}_{S}$ changes the sign of the spin quantization axis,
thus being the nonrelativistic analogue of time reversal. The remaining
operations $\mathcal{P}$, $\mathcal{M}$ and $\mathcal{C}_{2}$ are unitary
symmetries acting in the real space and these symmetries map each magnetic
sublattice onto itself. In contrast, the remaining four symmetries include the
$\mathcal{R}_{S}$ and map the two opposite magnetic sublattice on each other,
combined with the half unit cell translation $\boldsymbol{t}^{\prime}=\left(
\frac{1}{2}\frac{1}{2}\frac{1}{2}\right)  $. This reflects the fact that in
the nonmagnetic $Pnnm$ group $\mathcal{P},$ $\mathcal{M}_{z},$ $\mathcal{C}%
_{2z}$ are symmorphic operations, while the rest are glide planes and screw axes.

We can understand the anti-Kramers band structure by analyzing the action of
the symmetries in the momentum space. Let us now demonstrate the action of
$\mathcal{R}_{S}\mathcal{M}_{x}$:
\begin{equation}
\mathcal{R}_{S}\mathcal{M}_{x}E(k_{x},k_{y},k_{z},\sigma)=E(-k_{x},k_{y}%
,k_{z},-\sigma).\label{Mirror}%
\end{equation}
Thus, $E(k_{x},k_{y},k_{z},\uparrow)=E(-k_{x},k_{y},k_{z},\downarrow),$ and
for $k_{x}=0$ the two spins are degenerate, $E(0,k_{y},k_{z},\uparrow)=E(0,k_{y},k_{z},\downarrow).$ On the other hand, both spin-up and spin-down
bands structure must separately respect the Bloch theorem, so $E(\mathbf{k,}%
\uparrow)=E(\mathbf{k-G,}\downarrow).$ Using $\mathbf{G=(}2\pi,0,0),$ we see
that $E(-\pi,k_{y},k_{z},\uparrow)=E(\pi,k_{y},k_{z},\uparrow)=E(-\pi
,k_{y},k_{z},\downarrow).$ This set of rules (similarly for $k_{y}%
)\ $generates the set of the nodal planes shown in Fig. 1, and the functional
form of Eq. \ref{d}. Note that while the location of such nodal surfaces
depends on the exact nature of the operations in the second term in Eq.
\eqref{nonrelativ}, their presence is a universal feature of AK magnets (\textit{cf.},
\textit{e.g.}, previously discussed tetragonal RuO$_{2}$\cite{Libor}, MnO$_{2}$\cite{Noda2016} and
MnF$_{2}$\cite{Zunger}).

Away from the nodal planes the \textit{absolute value }of the spin splitting
obeys the full nonmagnetic crystal symmetry, while it \textit{sign}
alternates\cite{Reichlova2020}, which ensures that the material as a whole is
spin-compensated by symmetry. To illustrate this, we plot in Fig.3 bands along
the path $\Gamma RR^{\prime}\Gamma$ shown in Fig.~1(d). The full 
spin splitting and nodal structure can be seen in the Fermiology throughout
the BZ shown in Fig.~1(d) and in SI for all dopings. In SI we list the
degeneracies along the high symmetry lines and planes in the Brillouin zone in
Tables S1 and S2.

We emphasize the distinction of the spin splitting by antiferromagnetism and by relativistic 
interactions. The spin-orbit spin splitting
(\textit{e.g.,} Rashba or Ising in 2D) is characterized by spin-degenerate
points at the time-reversal symmetry invariant momenta protected by the
time-reversal symmetry. Away from these points, the bands spin-split
antisymmetrically,
\begin{equation}
E(\boldsymbol{k},\uparrow)=E(-\boldsymbol{k,\downarrow}),
\end{equation}
\textit{\ i.e}., the spin splitting $E(\boldsymbol{k},\uparrow
)=E(-\boldsymbol{k,\downarrow})$ follows a $p-,$ or $f-$wave rather than
$d-$wave, as in Eq. \ref{d}, symmetry. This, of course, is related to the fact
that spin orbit splitting can only happen in noncentrosymmetric crystals,
while in the absence of spin-orbit the $E(\boldsymbol{k},\sigma
)=E(-\boldsymbol{k,\sigma})$ holds independent of the magnetic structure (the
time reversal symmetry can be applied to the Schr\"{o}dinger equation for each
spin separately).

\begin{figure}[b]
\includegraphics[width = 1\linewidth]{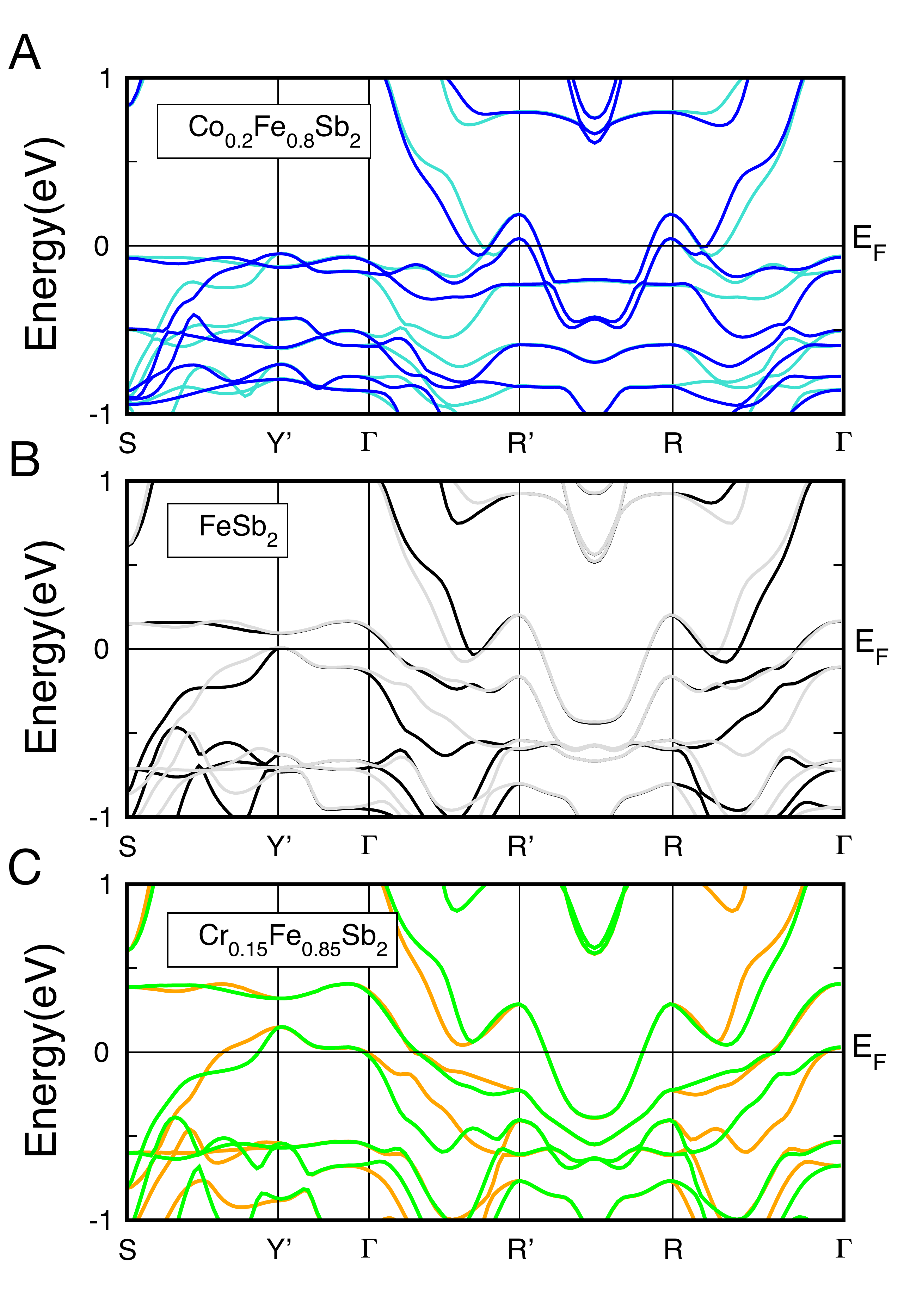}\caption{The nonrelativistic
band structure of FeSb$_{2}$ with the AK pattern, for electron doped (top),
stoichiometric (middle) and hole doped (bottom) and corresponding Fermi
surfaces. The "up" and "dn" bands split along low symmetric momenta in
accordance with Table S1 and S2. Furthermore, the
splitting exactly reverses along perpendicular momenta, i.e. $\Gamma R$ and
$\Gamma R^{\prime}$. This explains the overall net zero moment.}%
\label{BSFS}%
\end{figure}

\section{Relativistic symmetry analysis and electronic structure}

\begin{figure}[ptb]
\includegraphics[width = 0.95\linewidth] {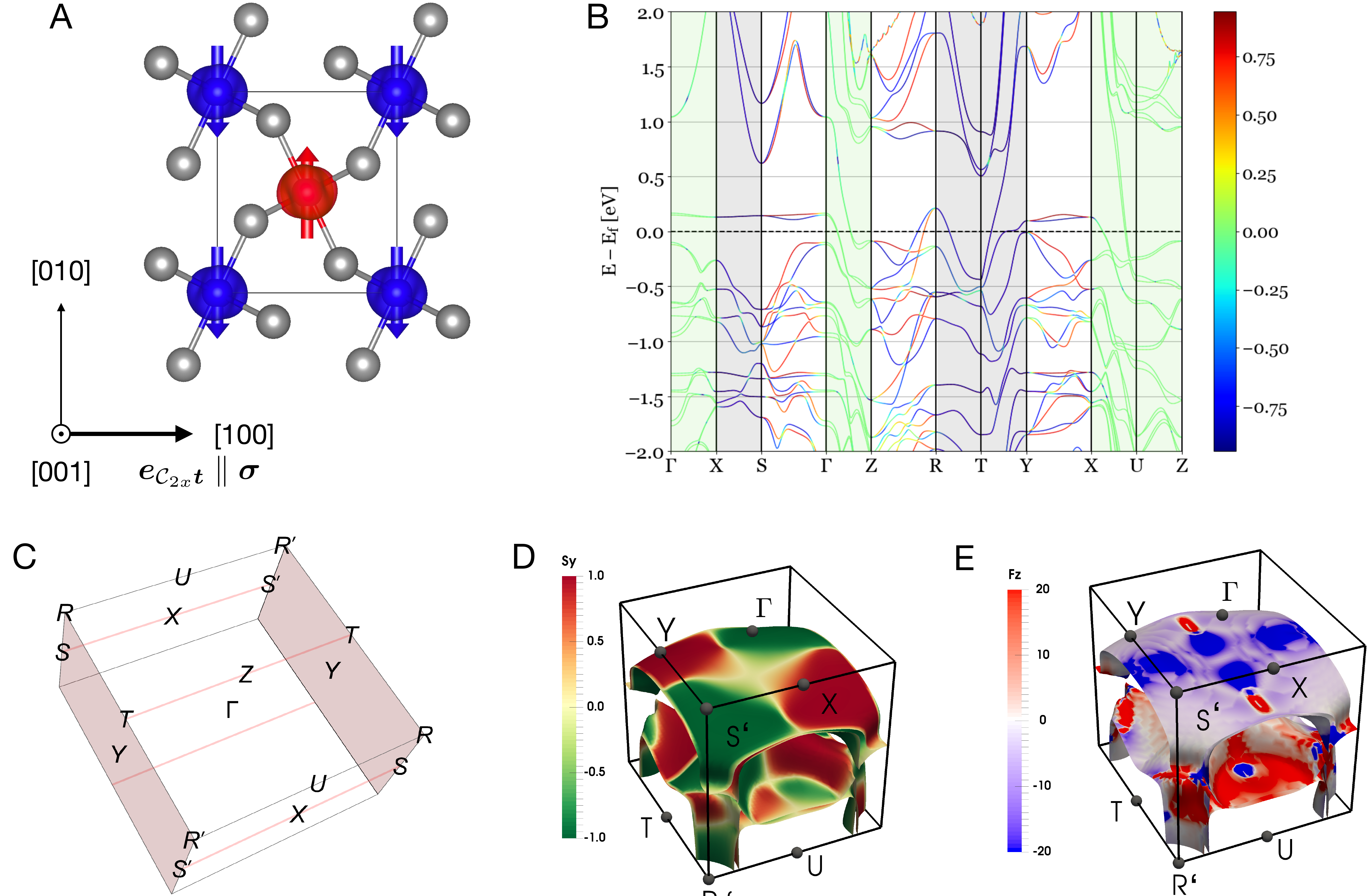}\caption{Electronic
structure of FeSb$_{2}$ with spin-orbit interaction. (a) Unit cell with the
calculated magnetization densities and marked symmetry group generators
$\mathcal{P},\mathcal{C}_{2a}\boldsymbol{t}$, and $\mathcal{C}_{2z}%
\mathcal{T}$, and Hall vector $\boldsymbol{\sigma}$. (b) Spin projected energy
bands calculated with spin-orbit interaction and sublattice magnetization along [010]
crystal direction. The grey shaded panels are spin degenerate, and green shaded ones are 
spin-split by spin-orbit, but still not spin polarized. (c) Brillouin zone band degeneracy manifolds in the
presence of spin-orbit interaction are marked by the red shading. (d) Spin
projected Fermi surfaces along [010] quantization axis. (e) Berry curvature
resolved on Fermi surface reveals large contributions from gapped AK nodal
features at $k_{x,y}=0$ planes marked by high-intensity blue and red colour.}%
\label{hall}%
\end{figure}

From the experimental point of view the most important prediction is that the
easy magnetization axis for FeSb$_{2}$ is calculated to be [010] (see Table
\ref{sigmatable}). Indeed, as mentioned in the introduction, while AK magnetism is, in
principle, not that rare,  those {so far} few identified as metals do not necessary have the magnetization direction conducive for
anomalous transport (as discussed, for instance, w.r.t. RuO$_{2}$ in Ref.
\cite{Libor}). In FeSb$_{2},$ to be specific, two out of three
orthorhombic directions manifest the latter ([100] and [010]), and the third
does not (Table~\ref{sigmatable}; see also Tables S1 and S2 in the SI for the
spin-orbit induced splittings). Fortunately, Cr (but not Co) doping not only
preserves the favorable magnetic anisotropy, but even enhances it. In
Fig.~\ref{hall}(a), we show the calculated magnetization density isosurfaces
for the magnetization along the [010] direction (calculated in VASP, see
Methods). The anisotropic spin densities highlight the breaking of the
$\boldsymbol{t}\mathcal{T}$ and $\mathcal{PT}$ symmetries.{\ }

\begin{table}[ptb]%
\begin{tabular}
[c]{lllllll}
& FeSb$_{2}$ & Cr$_{0.15}$Fe$_{0.85}$Sb$_{2}$ & Co$_{0.2}$Fe$_{0.8}$Sb$_{2}$ &
&  & \\\hline\hline
$\sigma_{yz}$ & 143 & -31 & 94 &  &  & \\\hline
next favorable axis & [001] & [001] & [010] &  &  & \\\hline
easy axis & [010] & [010] & [001] &  &  & \\\hline
MAE & -0.12 & -0.25 & -0.12 &  &  & \\\hline
\end{tabular}
\caption{The Hall conductivity in S/cm and the easy axis with magnetic
anisotropy energy in meV relative to the next favorable axis ([001]). }%
\label{sigmatable}%
\end{table}

In the following we will illustrate how the symmetry analysis of the spin
splitting proceeds in the relativistic case. In this case we need to take into
account the full magnetic space group (MSG), which depends on the selected
spin quantization axis; for instance, a mirror plane $\mathcal{M}_{y}$
conserves the spin components $s_{y},$ but flips the directions of $s_{x}$ and
$s_{z}.$ For the spin direction along the easy axis [010] direction the MSG is
$Pnn^{\prime}m^{\prime}$ (\textit{i.e.,} the glide and the mirror planes
symmetries for $y$ and $z$ planes, respectively, are preserved only when
combined with the time-reversal symmetry) and includes the following symmetry
operations ($\boldsymbol{t}^{\prime}=[\frac{1}{2}\frac{1}{2}\frac{1}{2}])$:%
\begin{equation}
\left\{  \mathcal{E},\mathcal{P}\right\}  +\boldsymbol{t}^{\prime}\left\{
\mathcal{C}_{2x},\mathcal{M}_{x}\right\}  +\mathcal{T}\left\{  \mathcal{C}%
_{2z},\mathcal{M}_{z}\right\}  +\boldsymbol{t}^{\prime}\mathcal{T}\left\{
\mathcal{C}_{2y},\mathcal{M}_{y}\right\}  .
\end{equation}
The notations are the same as in Eq.~\eqref{nonrelativ}, but the spatial and spin
space symmetries are now coupled, resulting in a different structure of the
symmetry group and band degeneracies. We also note that the $Pnn^{\prime
}m^{\prime}$ MSG describes a type-III antiferromagnet, that is, preserving the
nonmagnetic unit cell.

In Fig.~\ref{hall}(b) we show the calculated relativistic energy bands with
spins along [010] direction and including spin-orbit. By comparing these bands
with the nonrelativistic ones we confirm the collinear antiferromagnetism as
the main source of the spin-splitting in FeSb$_{2}$. The spin-orbit
interaction additionally splits certain high symmetry planes, lines and points
as summarized in Tables~S1, 2. However, the relativistic spin splitting is
much weaker.

Not all nodal planes shown in Fig. 1(d) are preserved, but one plane,
$k_{y}=\pm\pi,$ is. Let us explain why. Let us denote $\{\boldsymbol{t}%
^{\prime}|\mathcal{TC}_{2y}\}=\mathcal{TS}_{2y}$. {\ The space operation
alone, }$\mathcal{S}_{2y},$ is translating the first Fe, Fe1, into its AF
counterpart, Fe2. If the spin of Fe1(2) is $s_{y}(-s_{y}),$ $\mathcal{S}_{2y}
$ alone is not a symmetry operation, but  $\mathcal{TS}_{2y}.$ Conversely, for
the spin $s_{x}(-s_{x}),$ $\mathcal{S}_{2y}$ is a symmetry operation, but  {not
$\mathcal{TS}_{2y}.$} The same is true for $s_{z}.$ This is why for the spins
along $x,$ $y$ or $z,$ the MSG is, respectively, $Pn^{\prime}nm^{\prime
},Pnn^{\prime}m^{\prime}$ and $Pnnm.$

 We will now demonstrate the symmetry protection of nodal surfaces.
Let us now consider two Bloch functions, $\varphi_{\mathbf{k}}$
and $\varphi_{\mathbf{k}}^{\prime}$ related by the nonsymmorphic symmetry
operation $\mathcal{O}=\mathcal{TS}_{2y}:$%
\begin{align}
E_{\mathbf{k}^{\prime}}\varphi_{\mathbf{k}^{\prime}}^{\prime}(\mathbf{r)} &
\mathbf{=}H(\mathbf{r)}\varphi_{\mathbf{k}^{\prime}}^{\prime}(\mathbf{r)=}%
H(\mathbf{r)[}\mathcal{O}\varphi_{\mathbf{k}}(\mathbf{r)]}\\
\mathcal{O}[H(\mathbf{r)}\mathcal{\varphi}_{\mathbf{k}}(\mathbf{r)]} &
\mathbf{=}E_{\mathbf{k}}[\mathcal{O\varphi}_{\mathbf{k}}(\mathbf{r)]}%
\end{align}
Here we should remember that $\mathbf{k}^{\prime}=\mathcal{O}\mathbf{k\neq
k,}$ and that $\mathcal{O}$ is a symmetry operation of the Hamiltonian,
$\mathcal{O}H(\mathbf{r)}\mathcal{O}^{-1}=H(\mathbf{r).}$ We see that the
energies $E_{\mathbf{k}}$ and $E_{\mathbf{k}^{\prime}}$ of the two Bloch
states, $\varphi_{\mathbf{k}}(\mathbf{r)}$ and $\varphi_{\mathbf{k}^{\prime}%
}^{\prime}(\mathbf{r),}$ are identical. That means that either $\varphi
_{\mathbf{k}}(\mathbf{r)}$ and $\varphi_{\mathbf{k}^{\prime}}^{\prime
}(\mathbf{r)}$ are identical, or these are two genuinly degenerate states. In
order to demonstrate the latter, we calculate the overlap of the two
functions:
\begin{equation}
\left\langle \varphi_{\mathbf{k}^{\prime}}^{\prime}(\mathbf{r)|}%
\varphi_{\mathbf{k}}(\mathbf{r)}\right\rangle =\left\langle \mathcal{O}%
\varphi_{\mathbf{k}^{\prime}}^{\prime}(\mathbf{r)|}\mathcal{O}\varphi
_{\mathbf{k}}(\mathbf{r)}\right\rangle =\mathcal{O}^{2}\left\langle
\varphi_{\mathbf{k}}(\mathbf{r)|}\varphi_{\mathbf{k}}^{\prime}(\mathbf{r)}%
\right\rangle .
\nonumber
\end{equation}
Now, by definition, $\mathcal{O}^{2}${$=\boldsymbol{t}^{\prime}\mathcal{TC}%
_{2y}\boldsymbol{t}^{\prime}\mathcal{TC}_{2y}=e^{i(-k_{x}+k_{y}-k_{z}%
)/2}e^{i(k_{x}+k_{y}+k_{z})/2}=e^{ik_{y}}$ so }$\left\langle \varphi
_{\mathbf{k}}(\mathbf{r)|}\varphi_{\mathbf{k}^{\prime}}^{\prime}%
(\mathbf{r)}\right\rangle =e^{ik_{y}}\left\langle \varphi_{\mathbf{k}%
}(\mathbf{r)|}\varphi_{\mathbf{k}^{\prime}}^{\prime}(\mathbf{r)}\right\rangle
,$ that is to say, $\varphi_{\mathbf{k}}(\mathbf{r)}$ and $\varphi
_{\mathbf{k}^{\prime}}^{\prime}(\mathbf{r)}$ are not identical, moreover,
orthogonal, unless $k_{y}=0.$ Next, we note that two vectors, $k_{y}=\pi$ and
$k_{y}^{\prime}=-\pi,$ are related by the operation $\mathcal{O=TS}_{2y},$ and
at the same time by the reciprocal lattice vector $G=2\pi.$ This proves that
at the plane $k_{y}=\pi$ there are two orthogonal wave functions giving the
same electron energy, even with the spin-orbit, if the sublattice magnetization direction is $y.$
Obviously, the same proof applies to $k_{x}=\pi$ and the sublattice magnetization direction $x$. 

We note that our protecting antiunitary screw symmetry is oriented along the antiferromagnetic 
sublattices and its orientation can be controlled by the magnetic quantization axis (along $x$ or $y$ direction).
 This contrasts the ferromagnetic nodal surfaces protected by antiunitary screw  axis present for magnetization perpendicular to the symmetry axis\cite{Wu2018NS}.

{\ In Fig.~4(d) we show the spin polarization at the Fermi surface. We observe
that most of the Fermi sheets are spin-polarized, but near the former nodal
planes $k_{x,y}=0$ the spin direction continuously rotates from one direction
to the opposite. This feature makes the FeSb$_{2}$ antiferromagnet a promising
spin current generator \cite{Gonzalez-Hernandez2021}.}

Note that for the spins along [001] crystal axis, when the MSG is $Pnnm,$ the
three mirror symmetries prohibit the existence of a ferromagnetic pseudovector
and thus spontaneous Hall conductivity \cite{Libor}. For spins along [100] the
MSG allows a non-zero anomalous Hall conductivity and magnetooptical
(nondiagonal) components of the dialectric function, which are $\sigma_{xz}$
($\sigma_{yz}$) for $Pn^{\prime}nm^{\prime}$ ($Pnn^{\prime}m^{\prime}).$

\section{Anomalous electric and magnetooptical response}

\onecolumngrid

\twocolumngrid

\textbf{Berry curvature.} There are non-trivial ramifications in the anomalous
response of the momentum space-dependent splitting and presence of AK nodal
surfaces. We start by discussing the Berry curvature calculations shown in
Fig.~4(e) and Fig.~S2. We observe that the large Berry curvature originates
from the regions of the Brillouin zone which are degenerate without spin-orbit
interaction, The presence of the spin-orbit interaction splits these nodal
features and generates anisotropic Berry curvature as we show in Fig.~4(e),
including the former nodal planes. The fact that the AKAF nodal manifolds are
surfaces, and not lines, ensures a large Berry curvature contribution from
regions in the Brillouin zone where the nodal surface intersects the mirror planes.

\textbf{Crystal Hall effect.} The Hall vector direction follows simple
rules\cite{Libor}. The Hall vector $\boldsymbol{\sigma}$ is parallel to the
rotational axis of $\mathcal{C}_{2x}\boldsymbol{t}^{\prime}$ symmetry as shown
in Fig.~4(a). The amplitude of the intrinsic spontaneous Hall effect can be
calculated by integrating the Berry curvature over the Brillouin zone. We see
in Fig.~5(a) that while the Berry curvature changes sign, the cancellation is
incomplete the Hall conductivity $\sigma_{yz}$ is nonzero, while $\sigma_{xy}$
and $\sigma_{xz}$ vanish, in agreement with our symmetry analysis
\cite{Libor}. We show the resulting energy dependence of the Hall conductivity
in Fig.~5(a) for the three doping levels discussed in Section II (see also Fig.~S3).

\begin{figure}[ptb]
\includegraphics[width = 1\linewidth] {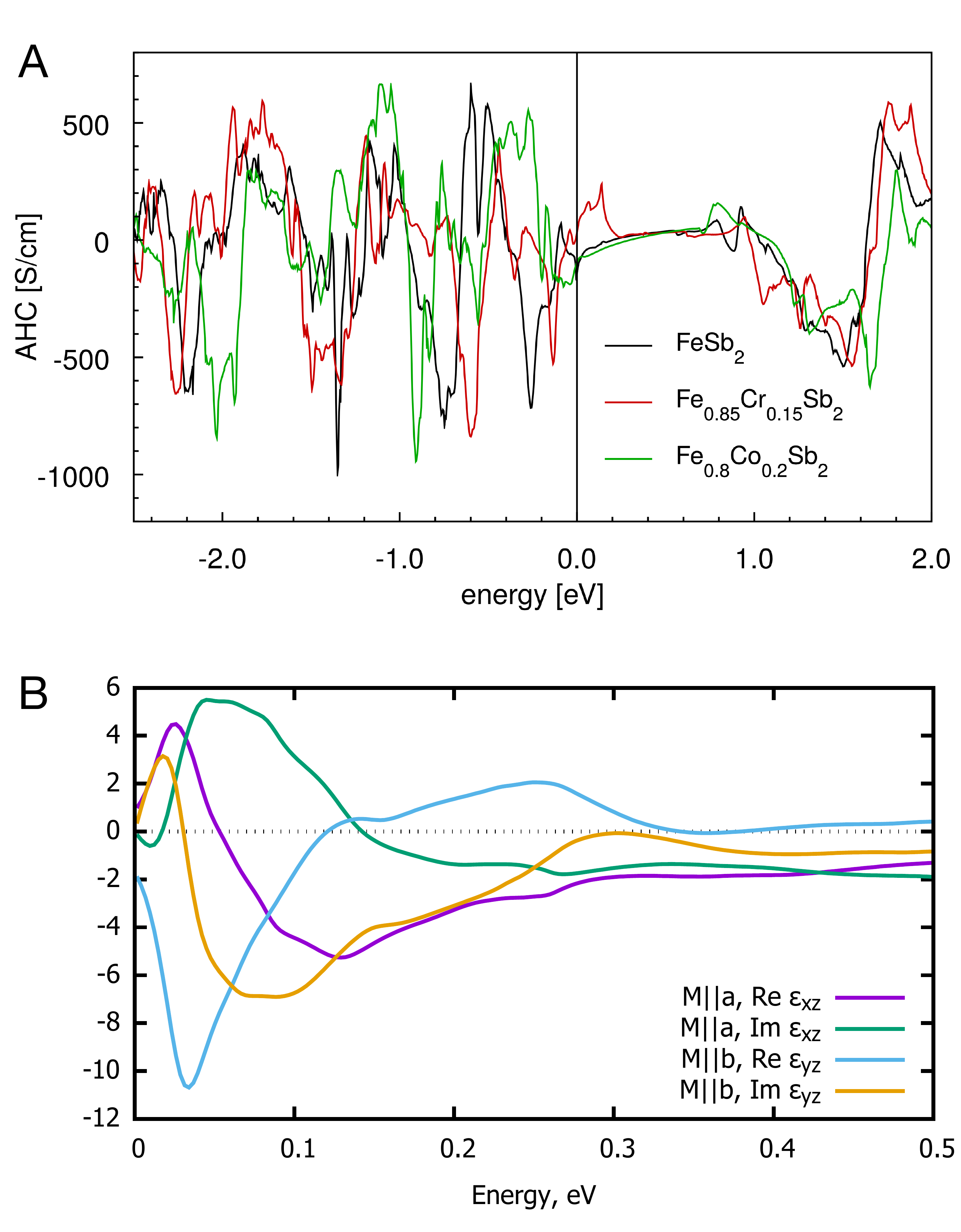} \label{Kerr}%
\caption{Anomalous charge and optical response. (a) Energy dependence of the
crystal Hall conductivity on the energy for the stoichiometric FeSb$_{2}$, and
electron and hole doped variants. (b) Energy (frequency) dependence of the
off-diagonal optical conductivity tensor calculated in stoichiometric
FeSb$_{2}$, which determines the 
magnetooptical Kerr effect\cite{Samanta2020}.}
\end{figure}

At the Fermi level, we obtain, for the [010] spin direction, 143~S/cm for the
undoped compound and 94~S/cm and -31~S/cm for the Co- and Cr-doped compounds,
respectively. Note however that the Co-doped compound has calculated $[001]$
easy axis, which does no afford an anomalous response. We conclude that the
Cr-doped FeSb$_{2}$ is the best candidate for the experimental observation of
the spontaneous crystal Hall effect. Previously, the crystal Hall effect was
experimentally observed in RuO$_{2}$ , but only after {a spin-reorientation
transition}, induced by a strong
external magnetic field \cite{Feng2020a}.

{\ \textbf{Crystal magnetooptical Kerr effect.} The magnetooptical Kerr effect
(MOKE) is a dynamic counterpart of the direct-current anomalous Hall effect
and follows the {analogous} symmetry-determined selection rules. It is a method of
choice for checking for ferro- and ferrimagnetic transition and is routinely
used as a litmus test for nonzero net magnetization and/or
noncollinear antiferromagnets\cite{Feng2015,Higo2018c,Feng2020b,Sivadas2016}.
In accordance to that, the }easy axis that affords a sizable Hall conductivity
also suggests, contrary to the common wisdom, presence of magnetooptical
effects in this fully compensated collinear material.

The complex Kerr rotation angle is proportional, in this case, to the $yz$
component of the complex dielectric function, $\varepsilon_{yz}(\omega).$ The
proportionality coefficient depends on the diagonal (Drude) dielectric
function, and thus on the sample-dependent relaxation rate. For that reason,
we present in Fig.~5(b) only the corresponding nondiagonal components of the
interband dielectric function, $\text{Re}\varepsilon_{{ij}}(\omega)$ and
$\text{Im}\varepsilon_{{ij}}(\omega)$ 
calculated from first-principles (see Methods). No that the %
${\omega\rightarrow0}${\ limit of the intraband conductivity tensor, }$\sigma
_{ij}(\omega)=\frac{\omega}{4\pi}\operatorname{Im}${$\varepsilon_{{ij}}%
(\omega),$ is the anomalous Hall conductivity\cite{Feng2015}. In agreement
with the symmetry analysis and the Hall effect calculations, we find for the
spin along [010] ([100]) a nonzero }${\varepsilon}${$_{yz}(\omega)$
($\varepsilon_{xz}(\omega)$). We obtain the largest magnitude for the
$\sigma_{yz}(\omega)$ at }$\omega\approx0.8$ eV, which is $\sim5$ times larger
{than the dc Hall conductivity. The magnitude of the magnetetooptical effects
in FeSb}$_{2}$ is comparable to that in typical ferromagnets.

\textbf{Methods} We used the Vienna Ab-initio Simulation Package (VASP)
\cite{VASP1} with the PBE-GGA approximation to the exchange correlation
potential \cite{GGA} and PAW pseudopotentials \cite{PAW1} to fully relax
FeSb$_{2}$ in a variety of magnetic patterns. We found that not only was the
AFMe pattern the lowest in energy, but the resulting lattice parameters of
$a=$ 5.8379, $b=$ 6.5248, and $c=$ 3.1811 matched extremely well with the
measured parameters of $a=$ 5.8328, $b=$6.53758, and $c=$3.19730
\cite{structure}. %The calculated lattice parameters for the non-magnetic \textcolor{(X X X)}  and AFMo type orders \textcolor{(X X X)} were slightly less well-matched and those of the FM order \textcolor{(X X X)}  were signficantly worse matched. 

To verify the energy orderings of the
different magnetic orderings with the highest possible accuracy, we calculated
each one with the FPLO code \cite{FPLO},also with the PBE-GGA approximation,
but adding spin-orbit coupling via a fully relativistic
four component solution to the Dirac equation. We found that
the energy differences shifted by less than 0.2 meV due to the inclusion of
this term. To simulate doping, we used the Wien2k code \cite{Wien2k} (again
with PBE-GGA) and employed the Virtual Crystal Approximation (VCA) with the
structure held constant according to the measurements of Ref. \cite{structure}. 
This shifts the charge of the ion cores to an average of the two charged
species being simulated. This method allows electrons/holes to be added to the
system in an average, band-like manner, consistent with experimental
measurements that find a lack of local moments in favor of a fluctuating
itinerant magnetic state \cite{mag_Koyama, mag_Zaliznyak}. To check the
validity of the VCA, we simulated Fe$_{0.5}$Cr$_{0.5}$Sb$_{2}$ with real,
rather than virtual, Cr doping and calculated the magnetic energies of the
various magnetic patterns (see Fig.~2(a)).

In our VASP calculations of the bandstructure and magnetisation densities in Figure 4 we set the energy cut-off to 520 eV,
use the momentum mesh of 7x6x12, and we use the GGA potential with the
Vosko-Wilk-Nusair potential. Furthermore, we use the Wannier90 code\cite{Pizzi2020} to
construct the Wannier functions. We calculate the spontaneous Hall
conductivity by integrating the Berry curvature in Brillouin zone in the
WannierTools code \cite{Wu2017b}. To corroborate the results for the Hall conductivity and to extend them to the doped cases Cr$_{0.15}$Fe$_{0.85}$Sb$_{2}$ and Co$_{0.2}$Fe$_{0.8}$Sb$_{2}$ via VCA we repeated the calculation using maximally projected Wannier
functions (WFs)\cite{Eschrig09} as provided by FPLO (version 19.00) in full
relativistic mode with subsequent integration of the Berry curvature. We
projected onto a set of Fe $3d$ and Sb $5p$ orbitals, which generates WFs for
all bands in $[-6,6]$~eV with only tiny Wannier fit errors of about 3~meV. The
Berry curvature was calculated by the method of Ref.~\cite{Wan06} using only
the dominant term (which was checked to be a rather small error). The
integration of the Berry curvature using the Wannier model was performed with
a k-mesh of $301\times{}301\times{}602$ and $400\times{}400\times{}800$ points
for the undoped and doped compounds, respectively, which was checked to be
converged to a below 5 percent error for the Hall conductivity at the Fermi
level. Calculations of the optical response were performed with the Wien2k
code \cite{Wien2k}.

\textbf{Acknowledgements} LS acknowledges the EU FET Open RIA Grant No. 766566
and SPIN+X Grant (DFG SFB TRR 173). LS and RGH acknowledge the computing time
granted on the supercomputer Mogon at Johannes Gutenberg University Mainz
(hpc.uni-mainz.de). IM acknowledges support from the U.S. Department of Energy
through Grant No. DE-SC0021089. MJ was supported by ONR through the NRL basic research program. 

\bibliography{FeSb2-4}
\end{document}

\title{Supplementary information: \\ Anti-Kramers antiferromagnetism in doped FeSb$_{2}$}
\author{I. I. Mazin}
\affiliation{Department of Physics and Astronomy, George Mason University, Fairfax, USA}
\affiliation{Center for Quantum Science and Engineering, George Mason University, Fairfax, USA}
\author{K. Koepernik}
\affiliation{Institute for Theoretical Solid State Physics, IFW Dresden, 01069 Dresden, Germany}
\author{M.D. Johannes}                                               
\affiliation{Center for Computational Materials Science, Naval Research Laboratory, Washington D.C., USA}
\author{Rafael Gonz\'{a}lez-Hern\'{a}ndez} 
\affiliation{Grupo de Investigaci\'{o}n en F\'{\i}sica Aplicada, Departamento de F\'{i}sica,Universidad del Norte, Barranquilla, Colombia}
\affiliation{Institut f\"ur Physik, Johannes Gutenberg Universit\"at Mainz, 55128 Mainz, Germany}
\author{Libor \v{S}mejkal}
\affiliation{Institut f\"ur Physik, Johannes Gutenberg Universit\"at Mainz, 55128 Mainz, Germany}
\affiliation{Institute of Physics, Czech Academy of Sciences, Cukrovarnick\'a 10, 162 00, Praha 6, Czech Republic}

\date{\today}

%%\pacs{75.50.Ee,75.47.-m,85.80.Jm}

\maketitle\
%\section{Staggered spin-momentum interaction model}

\begin{figure}[h]
\hspace*{0cm}\epsfig{width=1\columnwidth,angle=0,file=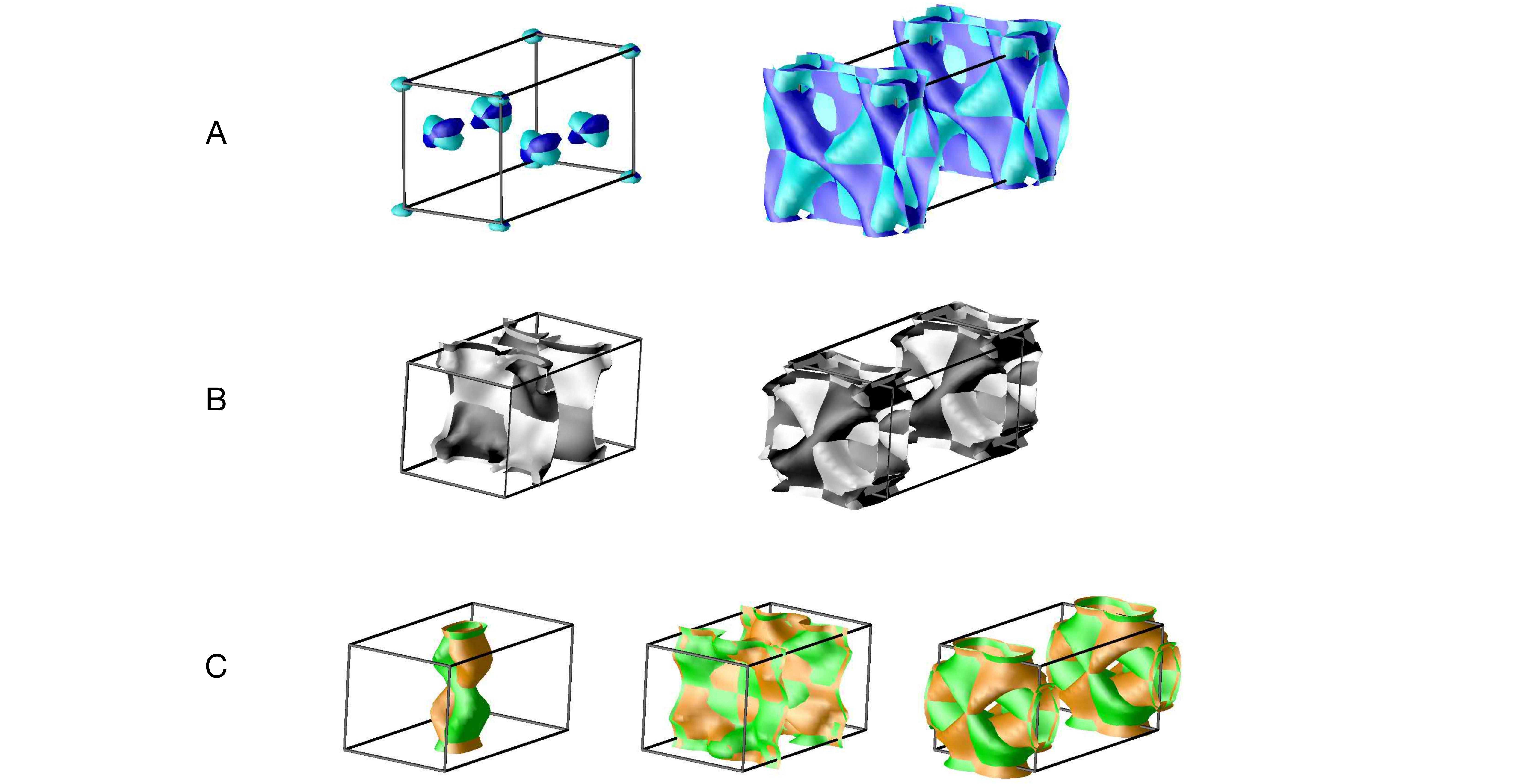}
\caption{Calculated Fermi surface pockets for: (a) Co$_{0.2}$Fe$_{0.8}$Sb$_{2}$, (b) FeSb$_{2}$, and (c) Cr$_{0.15}$Fe$_{0.85}$Sb$_{2}$.
The orientation of the Brillouin zone is shown in panel (d). Note that number of pockets increases with the number of electrons.}
\label{}
\end{figure}

\begin{figure}[h]
\hspace*{0cm}\epsfig{width=1\columnwidth,angle=0,file=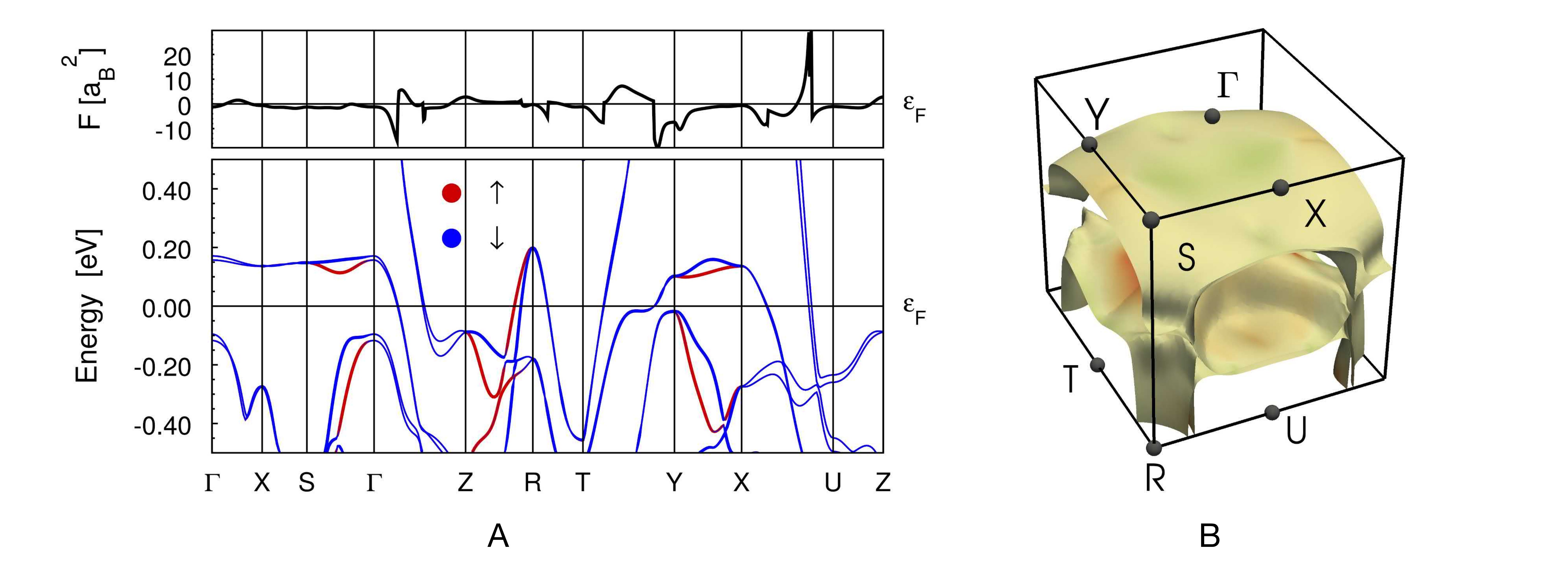}
\caption{(a) Calculated energy bands and Berry curvature $\mathcal{F}$ along high symmetry path in momentum space. (b) Spin polarization $S_{y}$ along the [010] direction plot on top of the Fermi surface  demonstrates vanishing net moment along the momenta with a large Berry curvature, cf. Fig.~4(e) in main text. The plots were calculated in the FPLO code, see Methods.\textcolor{red}{?????}}
\label{}
\end{figure}

\begin{figure}[h]
\hspace*{0cm}\epsfig{width=1\columnwidth,angle=0,file=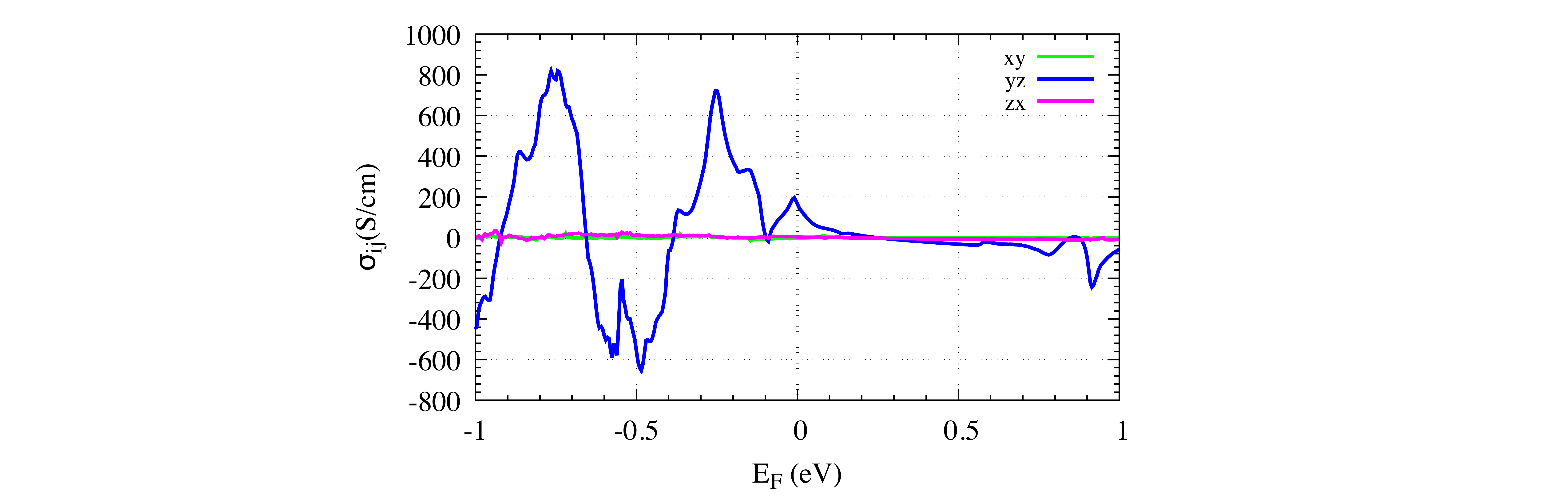}
\caption{Calculated spontanous intrinsic crystal Hall conductivity as a function of the Fermi energy, calculated using VASP and Wannier90 codes, for the spin direction [010]. Note that, in agreement with the symmetry analysis in the main text, the only nonzero component is $\sigma_{yz}$.}
\label{}
\end{figure}

\begin{table}[ptb]
\begin{tabular}{lllllllllllll}
%Field direction & \multicolumn{12}{c}{Lines} \\
\multicolumn{1}{r}{} & $\Gamma$X  & XS & SY & Y$\Gamma$ & $\Gamma$Z & XU & SR & YT & ZU & UR & RT & TZ \\  \hline  \hline
no SOC & \checkmark & \checkmark & \checkmark & \checkmark &\checkmark & \checkmark & \checkmark & \checkmark & \checkmark & \checkmark & \checkmark & \checkmark\\ \hline
100 & & \checkmark & \checkmark & \multicolumn{1}{r}{} & \multicolumn{1}{r}{} & \checkmark & \checkmark & \multicolumn{1}{r}{} & \checkmark & \checkmark & \multicolumn{1}{r}{} & \multicolumn{1}{r}{} \\ \hline
010 & \multicolumn{1}{r}{} & \checkmark & \checkmark & \multicolumn{1}{r}{} & \multicolumn{1}{r}{} & \multicolumn{1}{r}{} & \checkmark & \checkmark & \multicolumn{1}{r}{} & \multicolumn{1}{r}{} & \checkmark & \checkmark\\ \hline
001 & \checkmark & \checkmark & \checkmark & \checkmark & \checkmark & \multicolumn{1}{r}{} & \checkmark & \multicolumn{1}{r}{} & \multicolumn{1}{r}{} & \multicolumn{1}{r}{} & \multicolumn{1}{r}{} & \multicolumn{1}{r}{} \\ \hline
arbitrary & \multicolumn{1}{r}{} & \checkmark & \checkmark & \multicolumn{1}{r}{} & \multicolumn{1}{r}{} & \multicolumn{1}{r}{} & \multicolumn{1}{r}{} & \multicolumn{1}{r}{} & \multicolumn{1}{r}{} & \multicolumn{1}{r}{} & \multicolumn{1}{r}{} & \multicolumn{1}{r}{}
\end{tabular}
\caption{\label{symtable}
Spin degeneracy of high symmetry lines in the Brillouin zone without (line 2) and with (lines 3-6) spin-orbit interaction. The first column indicates the spin quantization axis for relativistic calculations.}
\end{table}

\begin{table}[]
\begin{tabular}{lllllll}
 & $\Gamma$XUZ & $\Gamma$YTZ & XSRU & SYTR & ZURT & ZURT \\ \hline\hline
no SOC &  &  \checkmark &  \checkmark &  \checkmark &  \checkmark &  \\ \hline
100 &  &  &  &  \checkmark &  &  \\  \hline
010 &  &  &  &  &  \checkmark &  \\   \hline
001 &  &  &  &  &  & 
\end{tabular}
\caption{\label{symtable2}
Same as in Tab.~\ref{symtable}, for high symmetry planes .} \end{table}